\documentclass[10pt]{iopart}

\usepackage{graphicx}

\begin{document}
\title[Ladder proof of nonlocality]{Ladder proof of nonlocality for two spin-half particles revisited}

\author{Jos\'{e} L Cereceda}
\address{C/Alto del Le\'{o}n 8, 4A, 28038 Madrid, Spain}

\ead{jl.cereceda@teleline.es}

\begin{abstract}
In this paper we extend the ladder proof of nonlocality without inequalities for two spin-half particles given by Boschi \etal (Boschi~D \etal 1997 \PRL\textbf{79} 2755) to the case in which the measurement settings of the apparatus measuring one of the particles are different from the measurement settings of the apparatus measuring the other particle. It is shown that, in any case, the proportion of particle pairs for which the contradiction with local realism goes through is maximized when the measurement settings are the same for each apparatus. Also we write down a Bell inequality for the experiment in question which is violated by quantum mechanics by an amount which is twice as much as the amount by which quantum mechanics violates the Bell inequality considered in the above paper by Boschi \textit{et al\/}.

\end{abstract}


\section{Introduction}

In 1993 Hardy gave a proof of Bell's theorem without inequalities for two spin-half particles (denoted by $A$ and $B$) and two alternative (noncommuting)observables per particle \cite{Hardy93}. Subsequently, Hardy \cite{Hardy97,Boschi} generalized this proof by considering an arbitrary number of observables per particle (from now on we will refer to this latter proof as the `ladder' proof). The consideration of more than two observables per particle results in an improvement of the percentage of particle pairs for which it is possible to obtain a contradiction with local realism---it grows from $9\%$ when only two observables are considered to almost $50\%$ which is obtained for a large number of observables and a state that is not quite maximally entangled. The ladder proof given in \cite{Hardy97,Boschi}, however, is not quite general. This is because the parameter (measurement setting) $c_k$, $k=0,1,\ldots, K$, defining the observable $A_k(c_k)$ for particle $A$, is the same as the parameter defining the corresponding observable $B_k(c_k)$ for particle $B$ (see equations (2)--(5) of \cite{Boschi}). In this paper we relax this restriction so that the observables $A_k$ and $B_k$ are initially specified by the independent parameters $\alpha_k$ and $\beta_k$, respectively. As we shall see, the resulting probability of getting a contradiction between the predictions of quantum mechanics and local realism will then explicitly depend on the choice of one of the observables involved, say $A_K$, through the corresponding parameter $\alpha_K$. This probability also depends on the number of settings (observables) considered and the quantum state describing the particles. We will show, however, that, for any given quantum state and any given $K$, the said probability is maximized whenever $\alpha_k =\beta_k$ for each $k$, that is, the probability is maximal when the setting defining the observable $A_k$ has the same value as the setting defining $B_k$. Finally, we will write down a Bell inequality (see equation (\ref{eq22})) for the experiment in question which is a generalization of the CHSH inequality \cite{CHSH} to the case of an arbitrary number of observables per particle \cite{Hardy91}. This inequality is violated by the relevant quantum predictions by an amount which is twice the amount by which quantum mechanics violates the Bell inequality considered in \cite{Boschi}.

\section{Extension of the original ladder proof}

Consider two spin-half particles in the entangled state
\begin{equation}
|\Psi\rangle = \alpha |+\rangle_A |+\rangle_B - \beta |-\rangle_A |-\rangle_B  \label{eq1}
\end{equation}
where $\{|+\rangle_A, |-\rangle_A\}$ ($\{|+\rangle_B, |-\rangle_B\}$) is an orthonormal basis in the state space of particle $A$ ($B$). Following \cite{Hardy97,Boschi}, it will be assumed that $\alpha$ and $\beta$ are taken to be real and positive, with $\alpha^2 + \beta^2 =1$. For each run of the experiment, on particle $A$ ($B$) we make a measurement of one two-valued ($\pm 1$) observable chosen from the set of $K+1$ observables $\{A_k\}$ ($\{B_k\}$), where $A_k$ and $B_k$ are defined as $A_k = |a_{k}^{+}\rangle \langle a_{k}^{+}| - |a_{k}^{-}\rangle \langle a_{k}^{-}|$ and $B_k = |b_{k}^{+}\rangle \langle b_{k}^{+}| - |b_{k}^{-}\rangle \langle b_{k}^{-}|$, and where the eigenvectors $|a_{k}^{\pm}\rangle$ and $|b_{k}^{\pm}\rangle$ are related to the original basis vectors $|\pm\rangle_A$ and $|\pm\rangle_B$ by
\begin{eqnarray}
|a_{k}^{+}\rangle =  \cos\alpha_k |+\rangle_A + \sin\alpha_k |-\rangle_A   \\
|a_{k}^{-}\rangle =  -\sin\alpha_k |+\rangle_A + \cos\alpha_k |-\rangle_A   \\
|b_{k}^{+}\rangle =  \cos\beta_k |+\rangle_B + \sin\beta_k |-\rangle_B   \\
|b_{k}^{-}\rangle =  -\sin\beta_k |+\rangle_B + \cos\beta_k |-\rangle_B .  
\end{eqnarray}

On the other hand, the observables $A_k$ and $B_k$ are required to satisfy the following conditions \cite{Hardy97,Boschi}:
\begin{eqnarray}
P_K = P(A_K=+1, B_K=+1) \neq 0  \label{eq6} \\
P(A_k =+1, B_{k-1}=-1) = 0 \qquad  \textrm{for $k=1$ to $K$} \label{eq7} \\
P(A_{k-1} =-1, B_k=+1) = 0 \qquad  \textrm{for $k=1$ to $K$} \label{eq8} \\
P(A_0 =+1, B_0=+1) = 0  \label{eq9}
\end{eqnarray}
where $P(A_k=m,B_{k^{\prime}}=n)$ denotes the joint probability that the outcome of the measurement of $A_k$ on particle $A$ is $m$, and that the outcome of the measurement of $B_{k^{\prime}}$ on particle $B$ is $n$, the pair of particles $A$ and $B$ being in the state (\ref{eq1}). The nonlocality argument based on equations (\ref{eq6})--(\ref{eq9}) is well known \cite{Hardy93,Hardy97,Boschi}, and it will not be repeated here. We merely note that the magnitude of the probability $P_K$ appearing in (\ref{eq6}) gives the proportion of particle pairs for which the contradiction between quantum mechanics and local realism goes through, and then it can be regarded as a direct measure of the degree of `nonlocality' inherent in such equations (\ref{eq6})--(\ref{eq9}).

It is readily shown that the fulfillment of the conditions in (\ref{eq7})--(\ref{eq9}) is equivalent to the fulfilment of the following:
\begin{eqnarray}
\tan\alpha_k / \tan\beta_{k-1} = - \alpha / \beta  \qquad \textrm{for $k=1$ to $K$} \label{eq10} \\
\tan\beta_k / \tan\alpha_{k-1} = - \alpha / \beta  \qquad \textrm{for $k=1$ to $K$} \label{eq11} \\
\tan\alpha_0 \tan\beta_0 = \alpha / \beta .  \label{eq12}
\end{eqnarray}
For a given quantum state (that is, for a given value of the ratio $\alpha/\beta$), equations (\ref{eq10})--(\ref{eq12}) contain $2K+2$ independent variables ($\alpha_k,\beta_k$) and $2K+1$ conditions. A glance at (\ref{eq10})--(\ref{eq12}) tells us that they determine all but one of the $2K+2$ variables, that is, once the value of one of the $2K+2$ variables is given (call this particular variable the free variable), then the remaining $2K+1$ variables get automatically fixed. Multiplying all $2K+1$ conditions in (\ref{eq10})--(\ref{eq12}), we obtain the constraint
\begin{equation}
\tan\alpha_K \tan\beta_K = \left(\alpha/\beta\right)^{2K+1} . \label{eq13}
\end{equation}
In what follows we choose $\alpha_K$ to be the free variable, so that the constraint in equation (\ref{eq13}) should be read as
\begin{equation}
\tan\beta_K = \left(\alpha/\beta\right)^{2K+1} \cot\alpha_K . \label{eq14}
\end{equation}
Now let us look at the probability $P_K$ in equation (\ref{eq6}). This probability is given by
\begin{equation}
\fl P_K =  \alpha^2  \cos^2 \alpha_K \cos^2 \beta_K  + \beta^2 \sin^2 \alpha_K \sin^2 \beta_K 
 -  \case{1}{2} \alpha\beta \sin 2\alpha_K \sin 2\beta_K . \label{eq15}
\end{equation}
But, as we have just seen, the variables $\alpha_K$ and $\beta_K$ are constrained to obey relation (\ref{eq14}). Thus, using (\ref{eq14}) in (\ref{eq15}) we obtain
\begin{equation}
P_K = \alpha^2 \left[ 1-\left(\alpha/\beta\right)^{2K} \right]^{2}
\frac{\cos^2 \alpha_K}{1+\left(\alpha/\beta\right)^{4K+2} \cot^2 \alpha_K}. \label{eq16}
\end{equation}
Equation (\ref{eq16}) is the most general expression for the probability in (\ref{eq6}) that is obtained when the conditions in (\ref{eq7})--(\ref{eq9}) are satisfied. Expression (\ref{eq16}) was already obtained elsewhere for the particular case in which $K=1$ \cite{Cereceda98}. From (\ref{eq16}) it is apparent that $P_K = 0$ whenever $\alpha = \beta$, that is, no contradiction with local realism arises for the maximally entangled state. It also vanishes for $\alpha=0$ or $\beta=0$ (i.e. for product states), as well as for $\alpha_K = n\pi/2$ ($n=0,\pm1,\pm2,\ldots\,$). Now we search for the value of $\tan^2 \alpha_K$ which, for fixed value of $\alpha/\beta$, maximizes the probability (\ref{eq16}). Recalling the trigonometric identity $\cos^2\alpha_K = \left(1+\tan^2\alpha_K \right)^{-1}$, and making the identification $x \equiv \tan^2\alpha_K$, this problem is equivalent to finding the value of $x$ that minimizes the function
\begin{equation}
f = 1+x+\left(\alpha/\beta\right)^{4K+2}\frac{1}{x}+\left(\alpha/\beta\right)^{4K+2} .  \label{eq17}
\end{equation}
From (\ref{eq17}) it is readily shown that $\partial^{2}f/\partial x^{2}$ remains finite and positive for all $\alpha_K \neq n\pi/2$. So, imposing the condition $\partial f/\partial x = 0$ we find that the value of $x\equiv \tan^2 \alpha_K$ that maximizes $P_K$ is
\begin{equation} 
\tan^2 \alpha_K = \left(\alpha/\beta\right)^{2K+1} .  \label{eq18}
\end{equation}

Comparing equations (\ref{eq18}) and (\ref{eq13}), we can see that the condition in (\ref{eq18}) is fulfilled if, and only if, $\tan\beta_K = \tan\alpha_K$. But, if $\tan\beta_K = \tan\alpha_K$, then the fulfilment of the conditions in (\ref{eq10}) and (\ref{eq11}) immediately implies that $\tan\beta_k = \tan\alpha_k$ for the remaining $k=0,1,\ldots, K-1$. This amounts to having $\beta_k = \alpha_k +n\pi$ for each $k$. Without loss of generality we may choose the solution $\beta_k = \alpha_k$. Thus we have shown that, as claimed, the probability (\ref{eq6}) is maximized when the measurement setting $\alpha_k$ of the apparatus at one end is equal to the corresponding measurement setting $\beta_k$ of the apparatus at the other end. The optimized probability is easily obtained by using (\ref{eq18}) in (\ref{eq16}). This gives
\begin{equation}
P_K = \left( \frac{\alpha\beta^{2K+1} - \beta\alpha^{2K+1}}
{\beta^{2K+1} + \alpha^{2K+1}} \right)^{2}  \label{eq19}
\end{equation}
which was originally derived by Hardy \cite{Hardy97,Boschi}. As shown in \cite{Hardy97,Boschi}, the maximum value of $P_K$ is $(50-\delta)\%$ which is realized for $K \to \infty$ and a state that is not quite maximally entangled.

\begin{figure}
\begin{center}
\includegraphics[width=3.2in]{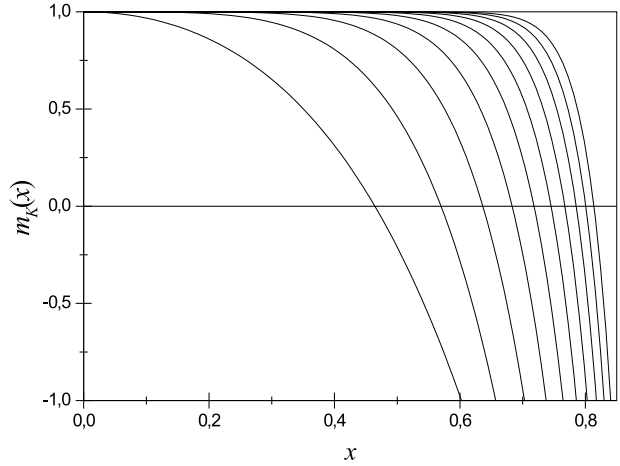}
\end{center}
\caption{\label{fig1}Plot of $m_K(x)$ for $K=1$ (leftmost curve) to $K=10$ (rightmost curve), and $0\leq x \leq 0.85$. The intersection point of the curve $m_K(x)$ with the central, horizontal axis determines the value of $r_1(K)$.}
\end{figure}

We note that the probability (\ref{eq19}) fulfils the symmetry property $P_K(\alpha,\beta)= P_K(\beta,\alpha)$. This means that if the value $r_1 = \alpha/\beta$ maximizes $P_K$ then the same holds true for the value $r_2 = \beta/\alpha$. It can be shown that the values $r_1$ and $r_2 =1/r_1$ which, for a given $K$, maximize (\ref{eq19}) correspond to the two nontrivial real roots of the polynomial
equation\footnote[1]{
The other (trivial) real root of the polynomial $m_K(x)$ is equal to $-1$ (with multiplicity 3). The remaining $4K-2$ roots of $m_K(x)$ are complex.}
\begin{eqnarray}
\fl m_K(x)=x^{4K+3} -(1+2K)x^{2K+3} -2Kx^{2K+2} - 2Kx^{2K+1} 
-(1+2K)x^{2K}+1=0. \label{eq20}  \nonumber   \\
\end{eqnarray}
Since $r_1,r_2 >0$ and $r_1r_2=1$ then necessarily one of $r_1$ or $r_2$ (say $r_1$) is less than 1, while the other is greater than 1 (note that we can never have $r_1=r_2=1$ because this corresponds to the maximally entangled state for which $P_K =0$). Equation (\ref{eq20}) can be solved numerically to obtain the roots $r_1$ and $r_2$ for any given value of $K$. It can be seen that, as $K \to \infty$, then $r_1 \to 1^{-}$ and $r_2 \to 1^{+}$, so that $r_2 - r_1 \to 0$. In \tref{table1} the values
\begin{table}
\caption{\label{table1}Numerical values of $r_1=\alpha/\beta$ and $r_2=\beta/\alpha$ giving the maximum probability $P_{K}^{\rm max}$ for the cases $K=1$ to $10$. $r_1$ and $r_2$ turn out to be the values of the (nontrivial) real zeros of the polynomial $m_K(x)$ (see \fref{fig1}).}
\begin{indented}
\item[]\begin{tabular}{@{}llllll}
\br
& $K=1$ & $K=2$ & $K=3$ & $K=4$ & $K=5$\\
\mr
$r_1$ & $0.464$ & $0.569$ & $0.636$ & $0.683$ & $0.718$\\
$r_2$ & $2.153$ & $1.754$ & $1.571$ & $1.463$ & $1.392$\\
$P_{K}^{\rm max}$ & $0.090$ & $0.174$ & $0.231$ & $0.270$ & $0.299$\\
\br
& $K=6$ & $K=7$ & $K=8$ & $K=9$ & $K=10$\\
\mr
$r_1$ & $0.745$ & $0.767$ & $0.785$ & $0.800$ & $0.813$ \\
$r_2$ & $1.341$ & $1.303$ & $1.273$ & $1.248$ & $1.229$ \\
$P_{K}^{\rm max}$ & $0.322$ & $0.339$ & $0.354$ & $0.365$ & $0.375$ \\
\br
\end{tabular}
\end{indented}
\end{table}
of $r_1$ and $r_2$, as well as the maximum probability $P_{K}^{\rm max}=P_K(r_1)=P_K(r_2)$, are listed for $K=1$ to $10$. Also, in \fref{fig1}, $m_K(x)$ (for $K=1$ to $10$) is represented
graphically for values of $x$ lying in the interval $[0,0.85]$.

\section{CHSH-type inequality for the ladder experiment}

So far we have tacitly assumed that the measurement apparatuses behave ideally. However, a real experiment testing the quantum predictions in (\ref{eq6})--(\ref{eq9}) does require the use of a Bell inequality because, in practice, it is not possible to attain the perfect correlations on which the ladder nonlocality contradiction relies. The Bell inequality used in the experiment reported in reference \cite{Boschi} is (see equation (20) of \cite{Boschi})
\begin{eqnarray}
\fl P(A_K=+1,B_K=+1) \leq P(A_0=+1,B_0=+1)  \nonumber  \\
+ \sum_{k=1}^{K} \, [P(A_k=+1,B_{k-1}=-1) +P(A_{k-1}=-1,B_k=+1)] \, .  \label{eq21}
\end{eqnarray}
In the ideal case, inequality (\ref{eq21}) is violated by quantum mechanics by an amount equal to $P_K$ since all terms on the right-hand side of (\ref{eq21}) vanish. In \cite{Boschi}, it is mentioned that inequality (\ref{eq21}) can be derived following the method described in \cite{Hardy91}. Incidentally, we would like to point out that, actually, the Bell inequality which is obtained by following the procedure in reference \cite{Hardy91} is not inequality (\ref{eq21}) but, instead, what is obtained is the related Bell inequality\footnote[2]{
It should be noted that, for the case $K=1$, inequality (\ref{eq22}) can be put equivalently as $P^{-}(A_0,B_0)+P^{+}(A_0,B_1)+P^{+}(A_1,B_0)+P^{+}(A_1,B_1) \leq 3$, which is a form of the CHSH inequality.}
\begin{equation}
\fl P^{+}(A_K,B_K) \leq P^{+}(A_0,B_0) 
+ \sum_{k=1}^{K} \, [P^{-}(A_k,B_{k-1}) + P^{-}(A_{k-1},B_k)]   \label{eq22}
\end{equation}
where
\begin{equation}
\fl P^{+}(A_k,B_{k^{\prime}}) = P(A_k=+1,B_{k^{\prime}}=+1) + P(A_k=-1,B_{k^{\prime}}=-1)  \label{eq23}
\end{equation}
and
\begin{equation}
\fl P^{-}(A_k,B_{k^{\prime}}) = P(A_k=+1,B_{k^{\prime}}=-1) + P(A_k=-1,B_{k^{\prime}}=+1). \label{eq24}
\end{equation}
We are going to show that, for the case in which the conditions in (\ref{eq7})--(\ref{eq9}) are satisfied, quantum mechanics violates inequality (\ref{eq22}) by an amount equal to $2P_K$. To see this, we first write (\ref{eq22}) in the equivalent form
\begin{equation}
S_K = P^{+}(A_K,B_K) - P^{+}(A_0,B_0) - 2\sum_{k=1}^{K} \, P^{-}(A_k,B_{k-1})\leq 0 \label{eq25}
\end{equation}
where we have used the fact that quantum mechanics predicts $P^{\pm}(A_k,B_{k^{\prime}}) = P^{\pm}(A_{k^{\prime}},B_k)$. We consider the case in which the settings $\alpha_k$ and $\beta_{k^{\prime}}$ are such that
\begin{eqnarray}
\tan\alpha_k = (-1)^{k} (\alpha/\beta)^{k+\frac{1}{2}}  \label{eq26}  \\
\tan\beta_{k^{\prime}} = (-1)^{k^{\prime}} (\alpha/\beta)^{k^{\prime}+\frac{1}{2}} \label{eq27}
\end{eqnarray}
(cf equation (25) of \cite{Boschi}). Relations (\ref{eq26}) and (\ref{eq27}) are obtained from (\ref{eq10})--(\ref{eq12}) by imposing that $\alpha_k = \beta_k$. For settings $\alpha_k$ and $\beta_{k^{\prime}}$ fulfilling (\ref{eq26}) and (\ref{eq27}) it can be shown that the quantum prediction for the sum of probabilities in (\ref{eq23}) and (\ref{eq24}) is given by
\pagebreak
\begin{equation}
P^{+}(A_k,B_{k^{\prime}}) =  \frac{1+x^{2(k+k^{\prime}+1)}-4\left( \frac{x}{1+x^2} \right) 
(-1)^{k+k^{\prime}} x^{k+k^{\prime}+1}} {(1+x^{2k+1})(1+x^{2k^{\prime}+1})}  \label{eq28}
\end{equation}
and 
\begin{equation}
P^{-}(A_k,B_{k^{\prime}}) = \frac{x^{2k+1}+x^{2k^{\prime}+1}+4\left( \frac{x}{1+x^2} \right)
(-1)^{k+k^{\prime}} x^{k+k^{\prime}+1}} {(1+x^{2k+1})(1+x^{2k^{\prime}+1})}   \label{eq29}
\end{equation}
where $k,k^{\prime}=0,1,\ldots,K$ and $x\equiv \alpha/\beta$. Of course we have $P^{+}(A_k,B_{k^{\prime}}) +P^{-}(A_k,B_{k^{\prime}}) =1$ and $P^{\pm}(A_k,B_{k^{\prime}}) = P^{\pm}(A_{k^{\prime}},B_k)$. Also we note that expressions (\ref{eq28}) and (\ref{eq29}) remain unchanged under the transformation $x \to 1/x$. Now, using (\ref{eq28}) and (\ref{eq29}), it can be seen that
\begin{eqnarray}
P^{+}(A_0,B_0) =  \frac{\left(1-x\right)^2}{1+x^2} \\
P^{+}(A_K,B_K) =  \frac{1+x^{4K+2}-\frac{4x^{2K+2}}{1+x^2}}{\left(1+x^{2K+1}\right)^2}
\end{eqnarray}
and
\begin{equation}
\sum_{k=1}^{K} \, P^{-}(A_k,B_{k-1}) = \frac{x(x-1)(x^{2K}-1)}{(1+x^2)(1+x^{2K+1})}
\end{equation}
so that
\begin{equation}
S_K = \frac{2x^2}{1+x^2}\left( \frac{1-x^{2K}}{1+x^{2K+1}} \right)^2 = 2P_K 
\end{equation}
(cf equation (\ref{eq19})). So, in the ideal case, the inequality $S_K \leq 0$ is violated by quantum mechanics by an amount $2P_K$, with the maximum violation $S_K \to 1$ occurring in the limit $K \to \infty$ and $x \to 1$. In this same limit, we have that $P^{+}(A_0,B_0) \to 0$, $P^{+}(A_K,B_K) \to 1$, and $P^{-}(A_k,B_{k-1})=P^{-}(A_{k-1},B_k) \to 0$ for $k=1$ to $K$.

It is worth noting that, in the limit considered, a direct contradiction between quantum mechanics and local realism emerges. Indeed, in order for a local hidden variable theory (LHV) to reproduce the quantum predictions in the above limit, the following $2K+2$ conditions must be met:
\begin{eqnarray*}
A_0 B_0 & = & -1  \\
A_1 B_0 & = & +1  \\
A_0 B_1 & = & +1  \\
\quad\ldots &  & \\
A_k B_{k-1} & = &+1  \\
A_{k-1} B_k & = &+1  \\
\quad\ldots  &  & \\
A_K B_K & = &+1
\end{eqnarray*}
where $A_k$ ($B_k$) denotes the result ($+1$ or $-1$) of the measurement of the observable $A_k$ ($B_k$). In the context of an LHV theory, the values of $A_k$ and $B_k$ can be regarded as elements of physical reality \cite{EPR} corresponding to the observables $A_k$ and $B_k$. However, the above $2K+2$ relations cannot be satisfied simultaneously since each quantity $A_k$ and $B_k$ appears twice on the left-hand side. Hence, the product of all these $2K+2$ relations must be equal to $+1$ on the left-hand side, but equal to $-1$ on the right-hand side. This result indicates that, in order to get a direct contradiction between quantum mechanics and local realism for two spin-half particles, it is necessary to consider an infinite number of observables per particle \cite{Hardy91}. By contrast, for systems consisting of three or more spin-half particles a direct contradiction can be obtained with only two observables per particle \cite{GHZ89,GHSZ90+M90}.

\section{Conclusion}

In summary, in this paper we have extended the ladder proof of nonlocality for two spin-half particles given in \cite{Hardy97,Boschi} to the case in which the settings of the measurement apparatuses are initially specified by the independent parameters $\alpha_k$ and $\beta_k$ ($k=0,1,\ldots,K$). We have shown that the proportion $P_K$ of particle pairs contradicting local realism is maximized whenever $\alpha_k = \beta_k$ for each $k$. It is this latter case which is considered in \cite{Hardy97,Boschi}. On the other hand, we have mathematically characterized the values $r_1=\alpha/\beta$ and $r_2=\beta/\alpha$ maximizing the probability (\ref{eq19}) obtained by Hardy as being the two nontrivial real roots of the polynomial $m_{K}(x)$ in equation (\ref{eq20}). Finally, we have written down a Bell inequality in equation (\ref{eq22}) for which, for the experiment in question, quantum mechanics predicts an amount of violation which is twice the amount by which quantum mechanics violates the Bell inequality (\ref{eq21}) considered in \cite{Boschi}, inequality (\ref{eq22}) being a generalization of the CHSH inequality to arbitrary many settings (observables) per particle.

\end{document}